\documentclass[journal]{IEEEtran}
\usepackage{amsmath} 
\usepackage{amssymb} 
\usepackage{epsfig} 
\usepackage{graphicx}
\usepackage{url} 
\usepackage{listings} 
\newtheorem{definition}{Definition}
\newtheorem{theorem}{Theorem}

\renewcommand{\theequation}{\arabic{equation}}
\renewcommand{\thetable}{\arabic{table}}
\begin{document}
\title{Preserving HTTP Sessions in Vehicular Environments}
\author{Yibei Ling,~\IEEEmembership{Senior~Member,~IEEE,} Wai Chen,~\IEEEmembership{Senior~Member,~IEEE,} 
T. Russell Hsing,~\IEEEmembership{Fellow,~IEEE,} and 
Onur Altintas,~\IEEEmembership{Member,~ACM,~IEEE}%
\thanks{Manuscript received: September 20, 2003; revised: November 06, 2005; accepted: October 13, 2006}% 
\thanks{Yibei Ling, Wai Chen, and T. Russell Hsing are with 
Applied Research, Telcordia Technologies, 
One Telcordia Drive, Piscataway, NJ 08854,  USA} 
\thanks{Emails: \{lingy,wchen,trh\}@research.telcordia.com}
\thanks{Onur Altintas is with Research 
\& Developement Division, Toyota InfoTechnology Center, Co., Ltd., 
6-6-20 Akasaka, Minato-ku, Tokyo, 107-0052 Japan. 
This work was done when the author was 
at Toyota InfoTechnology Center, USA and also a visiting 
researcher at Telcordia Technologies}
\thanks{Email: onur@jp.toyota-itc.com}}

\markboth{IEEE Transactions on Vehicular Technology}{Shell \MakeLowercase{\textit{et al.}}: Bare Demo of IEEEtran.cls for Journals}
\maketitle

\begin{abstract}
Wireless Internet in the in-vehicle environment 
is an evolving reality that reflects the gradual maturity
of wireless technologies. Its 
complexity is reflected in 
the diversity of wireless technologies and dynamically changing 
network environments.
The ability to adapt to the dynamics of such environments
and to survive transient failures due to network handoffs
are fundamentally important in failure-prone vehicular environments.   
In this paper we identify several new issues arising from network heterogeneity 
in vehicular environments and 
concentrate on designing and implementing a network-aware 
prototype system that supports HTTP session continuity in 
the presence of 
network volatility, with the emphasis on the following specifically
tailored features:   
(1) automatic and transparent HTTP failure recovery, 
(2) network awareness and adaptation, 
(3) application-layer preemptive network handoff.  
Experimental results gathered from real application environments
based on CDMA {\it 1xRTT} and IEEE 802 networks are presented
and analyzed.  
\end{abstract}
\IEEEpeerreviewmaketitle
\section{Introduction}
\PARstart{R}{apid} technology advancement in mobile communication 
has significantly changed our
way of communication and information exchange. 
The proliferation of various 
portable devices such as laptop computers, 
personal digital assistants (PDA), 
cellular phones, and the ubiquity of 
various cellular networks such as CDMA, CDPD, GSM, and 
wireless LAN (WLAN) IEEE 802.11a/b/g, offer a 
powerful mobile computing platform that
completely 
frees devices/computers from being tethered to the wired line. 
Thus convergence and integration of 
Internet \cite{Claffy1998} and wireless technologies
spark new ways of using Internet
and spurs a growing demand for a wide range of 
telematics applications. 

Ensuring HTTP session continuity in vehicular environments
is of paramount importance to vehicular Internet access, 
since network connectivity is inherently unreliable due to 
the presence of network handoff and the 
existence of blind coverage spots. 
However, the issue of HTTP session continuity 
has been largely overlooked: 
it appears to be irrelevant to real applications since driving and web 
surfing at the same time are generally considered as 
a risk distraction.  Such a viewpoint   
is reflected in the comment  ``don't walk while telnet'ing" by 
Henning Schulzrinne \cite{Henning2000}.  

The emerging telematics applications, especially
a wide range of rear-seat applications including HTTP-based application \cite{Claffy1998} 
highlight the importance of HTTP session continuity.   
However, the gap between the promise of in-vehicle Internet access 
and reality remains wide,  mainly reflecting in the following 
distinct aspects: 
\begin{enumerate}
\item the presence of multiple networks with 
different characteristics in 
overlapping coverage areas;
\item the inherently 
unreliable network connectivity in vehicular environments 
due to the presence of blind coverage spots, network handoffs and 
high vehicle mobility; 
\item existing applications are essentially network-oblivious and  
lack the ability to cope with dynamically changing
network environments.
\end{enumerate}
 
In a vehicular environment, 
a mobile host (MH hereafter) constantly changes its geographic location, 
resulting in the switching of coverage responsibility from one 
base station to another.
Such network switch, called network handoff,  
causes the MH to be temporarily 
disconnected to the network, thus resulting in a disruption of on-going sessions. 
The adverse effect of handoff upon system performance could become 
prominent in highly mobile vehicular environments \cite{Anne1999}.
Thus, network handoff is a major mobility issue that 
has been receiving a great deal of 
research interest 
in both industry and academia.

From the perspective of a MH, 
the network handoff can be classified into two distinct categories: 
(1) horizontal handoff and (2) vertical handoff \cite{stemm98vertical}.
A horizontal handoff refers to a network switch taking place in 
homogeneous network environments. For instance,    
a handoff occurs during subnet-crossing in a WLAN network.
A vertical handoff refers to a network switch taking place 
in heterogeneous network environments, 
such as a handoff between a CDMA cellular network and a Wireless LAN (WLAN) network.
 
The vehicular Internet access is more prone to transient 
network failures than wired line Internet access mainly for two reasons. First, 
handoffs induced by vehicle mobility are far more common than pedestrian mobility due to 
high vehicle mobility. Second,
coverage of wireless network is far from ubiquitous: the presence of 
blind coverage spots becomes a source of losing network connection in vehicular environments.  
Another equally important aspect is the efficiency of HTTP failure recovery, especially for 
long HTTP session (downloading large files) in relatively low bandwidth
wireless network environments. 
In a vehicular environment, 
when an HTTP session is punctuated by transient network failures,  
the HTTP session (a kind of TCP session) needs to be
restarted from scratch. This could incur expensive network resource in 
wireless vehicular environments since transient 
network failures could be frequently 
encountered events.    
Therefore,  an efficient and transparent HTTP failure recovery
is an essential feature required in failure-prone vehicular environments.

Wireless vehicular environments are inherently heterogeneous, 
consisting of different types of wireless technologies with
different characteristics. 
It is well known that 
the CDMA and WLAN networks
differ significantly in terms of bandwidth capacity and coverage range:
the CDMA network is of relative low bandwidth capacity but has a wide coverage area, while 
the WLAN network is of high bandwidth capacity but has a 
narrow coverage
area \cite{Cheshire1995,Baker1996,Ling2004,Ylianttila2001}. 

The coexistence of CDMA and WLAN networks in 
overlapping coverage areas
gives rise to a fundamental phenomenon in 
heterogeneous network environments called 
{\it network hysteresis} (access history dependency), which 
refers to the inherent tendency of an established
TCP (HTTP) session to stay with a network once the 
session has been established over the network.  

This situation can be exemplified as follows:
a MH with multiple network interfaces initiates a 
(TCP or HTTP) session over a low-bandwidth network such as 
the CDMA network, the MH then moves into the overlapping coverage area 
with a high-bandwidth network such as WLAN network. 
Despite the presence of the high-bandwidth network,
the MH continues to stay with the low-bandwidth network 
(CDMA). It is  because that  
(1) the CDMA network has a wider coverage area as compared to the WLAN network, 
(2) there is an inherent tendency to maintain connectivity 
over a network after the initial (TCP or HTTP) session has been
established 
(inherent network environment obliviousness).

The phenomenon of the {\it network hysteresis}, which arises only  
in heterogeneous wireless network environments,
carries an adverse performance implication for 
the in-vehicular applications: 
it prevents the MH from utilizing 
the best available network resource, 
thereby impairing system performance in 
inherently heterogeneous vehicular network environments. 
To our best knowledge, 
the impact of network hysteresis on system performance
has not been considered before.  

In this paper we introduce the preemptive handoff to address 
the problems arising from the network hysteresis. 
The main idea behind the preemptive handoff 
is to dynamically reselect the best available network interface 
in the presence of 
multiple network interfaces. This involves 
automatically relinquishing  
a lower-bandwidth connection and reestablishing a higher-bandwidth 
connection while at the same time maintaining session continuity.
Preemptive handoff in spirit is similar to 
the downward vertical handoff which is defined as 
network handoff from a low-bandwidth network to a high-bandwidth network \cite{stemm98vertical}.  
However, the proposed preemptive handoff differs markedly from   
vertical or horizontal handoff 
in its goal to optimize system throughput, in addition to ensuring session continuity 
during handoffs. 
In this paper we present an architectural framework as well as a
prototype system based on this framework, stressing particularly on
the following aspects. 
\begin{itemize}
\item supporting HTTP session continuity in the presence of handoffs;
\item network awareness and network adaptation;
\item carrier-independence; 
\item automatic and efficient HTTP failure recovery; 
\item access transparence.
\end{itemize}

The remainder of this paper is organized as follows:
Section 2 gives a brief review of  related mobility approaches in the literature, 
with the focus on mobile IP and Session Initiation Protocol (SIP hereafter).
Section 3 presents a system architecture, as well as its implementation,
for HTTP failure recovery.  Section 4 
describes our heterogeneous network access testbed 
with cellular 
CDPD and CDMA networks, and WLAN networks. The experimental results
are presented and discussed.  
Section 5 concludes the paper with possible future extensions.

\section{Overview of Related Work}
In this section we first introduce the notions of HTTP session, 
and session-level, and packet-level HTTP failure recovery, 
followed by a review of  
Mobile IP (MIP hereafter) and 
Session Initiation Protocol (SIP hereafter) .
We then discuss the inadequacy of SIP and MIP 
in dealing with HTTP session in a practical setting.  
We also briefly discuss the Wireless Application Protocol (WAP) 
and touch on the Tarantella software system. 
Finally, we look beyond MIP and SIP for solution to problems 
stemming from network 
heterogeneity. \\

\begin{definition}
An HTTP session consists of
an HTTP request and the
HTTP response to that request.  \hfill $\blacktriangle$  \\
\end{definition} 

An HTTP request includes both
explicit and implicit HTTP request. 
An explicit HTTP request is initiated manually, whereas 
an implicit HTTP request, as an ancillary event triggered by 
an explicit HTTP request, is initiated transparently by 
the Web client. 
To put more intuitively, an HTTP session is started with a 
request initiated from 
a Web client to a Web server and terminated with the 
reception of the entire response from 
the origin Web server. The following definitions are 
given to distinguish the two granular levels of HTTP failure recovery. \\
\begin{definition}
A session-level HTTP failure recovery is atomic or indivisible
with respect to a given point of attachment. \hfill $\blacktriangle$ \\
\end{definition} 

Session-level HTTP failure recovery carries an all-or-nothing implication, 
meaning that 
an HTTP failure cannot be partially recovered. 
For instance, downloading an HTTP file either completes or fails with
respect to a given point of attachment (network).
Session-level HTTP failure recovery could be
relatively palatable in a wired network environment because 
of relatively high bandwidth. However, it 
could become extremely
annoying when network handoffs 
occur in the midst of long-lived HTTP sessions, 
especially when most of the transfer has taken place.
Therefore, the weakness of session-level HTTP failure recovery 
is its inefficiency of network utilization. A
failure recovery requires that file transferring be
restarted from scratch, which could be very 
costly in relatively low bandwidth network environments. 
In addition, a frequent handoff in failure-prone vehicular environments could 
result in a frequent failure recovery, which is likely to form
an endless cycle where failures and premature recoveries 
are interlocked.
As a result, session-level HTTP failure recovery by nature 
cannot guarantee the progressiveness in the presence of frequent handoff.   
Packet-level HTTP failure recovery is thus proposed to address the 
deficiency of its session-level counterpart. Its definition is  given as  below: \\

\begin{definition}
A packet-level HTTP failure recovery is divisible 
with respect to the point of
attachment.  \hfill $\blacktriangle$  \\
\end{definition} 

Packet-level HTTP failure recovery differs from the session-level counterpart in
its ability to avoid HTTP session being restarted 
from scratch after 
a failure occurs. This means that constituent data packets obtained from different 
points of attachments (networks) can be seamlessly pieced together.   
The packet-level HTTP failure recovery improves 
upon the session-level counterpart
in its failure-recovery efficiency, 
thus making it more suitable in failure-prone vehicular environments.  
Now we are in a position to review existing approaches in 
the literature, 
and evaluate their pros and cons in a practical setting. 

Mobile IP (MIP) is a standard network-layer mobility protocol,  allowing 
a MH to maintain session continuity when roaming to a different network
\cite{Fikouras1999,Anne1999,Katayama2001,karagiannis-mobility}.  
It is implemented via IP-in-IP encapsulation, IP 
tunneling, and IP decapsulation.
Under the MIP scheme, the MH 
consists of (1) a fixed IP (primary IP or home address); 
(2) a care-of address that is changed with 
the change of point of attachment. 
When the MH moves to a new location (foreign network), it
registers the IP address of a foreign network with the home agent 
located on the user's home network. The home agent is
thereby able to transparently tunnel all packets to the user's current location 
via the IP-in-IP encapsulation. Upon receipt of data
packets from the home agent, the foreign agent 
decapsulates packets and delivers them to the MH. 

The most attractive feature of MIP is its    
application transparency. It supports mobility and session continuity
without the need to modify existing applications.
However, the benefits of the MIP come at a performance price: 
MIP requires performing run-time IP-in-IP  encapsulation and decapsulation for each 
IP packet, and introduces 
triangle routing as well as an additional 
delay between home agent and foreign agent.  
As a result, mobile IP could increase the access latency by $45$\%  
in a typical campus environment \cite{Weldlund99}.
Much research effort has focused on routing optimization in order to 
minimize the effect of costly triangle routing
\cite{karagiannis-mobility,Parsa1999,Anjum1999,Caceres1995,Stathes2002,Perkins98,Perkins99},
and reduce the handoff time \cite{stemm98vertical,Ylianttila2001}. 
  
Session Initiation Protocol (SIP) is widely used as the 
replacement of H.323 protocol
for multimedia streaming, 
various UDP-based applications such as 
Internet conferencing, telephony, instant messaging 
and real-time event notification.
Major components in SIP are SIP user agent and SIP proxy/redirect servers. 
Under the SIP scheme,  a SIP user agent residing in the MH is responsible for 
updating the home SIP proxy server with the current location (IP address), 
and managing the SIP-aware applications.  
Each SIP user is addressed by a unique SIP identity (an email-like address), which is
initially registered in a SIP registrar managed by the SIP proxy server, together with 
the location (IP address) of the MH. To communicate with a peer MH,  the MH 
sends an INVITE message with the SIP identifier of the peer MH 
to the SIP proxy server, which in turn returns the current location (IP address) 
of the peer MH via dynamic binding (identifying the IP address through its SIP identifier).
The MH can then use this IP address to directly send an INVITE message to the peer MH.  
Since it is being implemented  on top of UDP or TCP, SIP is widely used 
as a signaling protocol for session establishment and management in 
mobile environments for terminal mobility and 
service mobility. 
It can also be used to support streaming-based session (VoIP) 
continuity such as RTP.
However, it does not work very well for TCP-based applications 
\cite{Henning2000, Weldlund99}.
\begin{figure}[htb]
\centering
\centerline{\psfig{file=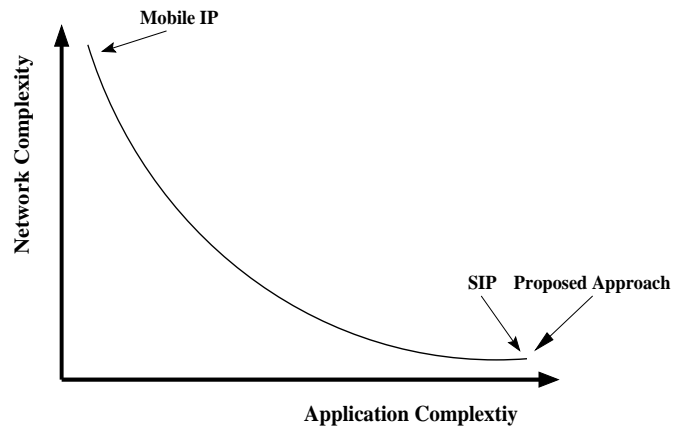,height=2.2in,width=3.5in}}
\caption{Network Complexity vs. Application Complexity of Mobile IP and SIP}
\label{fig:complex} 
\end{figure}

MIP and  SIP are two distinct and competing paradigms for supporting mobility and
session continuity, representing two opposite extremes in terms of network 
and application complexities.
MIP has a high level of network complexity with a relatively low application complexity.
In contrast,  SIP has a relatively 
high level of application complexity  with a minimum level of network 
complexity, as shown in Fig({\ref{fig:complex}). \\

Market environment issues, however, are often  perceived
as the single most important challenge to mobile IP and SIP deployment 
because of the autonomy of service providers and wireless operators, 
which adds another level of complexity 
in vehicular Internet access. 
Deployment of Mobile IP and SIP proxy/redirect server   
is required to have a complete control over underlying wireless networks. 
This is only  possible for wireless operators such as AT\&T and Verizon 
\cite{stemm98vertical}. 
Thus it could be inoperative under many conditions for wireless service subscribers/customers.
For instance, a SIP user agent may be unable to receive a session 
initiation call from the 
SIP server managed by different service providers and wireless operators. 
The vertical handoff scheme proposed in \cite{stemm98vertical}
requires that a home agent be able to multicast data packets to a group of 
of base stations in which a foreign agent resides, which is constrained 
by a variety of factors in practice. For those networks that we cannot control, 
it is impossible to put a foreign agent  into their base stations.  
In reality, a vertical handoff normally could take more than $10$ seconds (it will be shown later on),
which could cause an unrecoverable TCP failure (MIP) because it 
exceeds TCP timeout \cite{Anne1999,Okoshi1999}.  

Tarantella is a commercial thin-client software system that 
enables enterprises to provide users anywhere with managed secure 
web-based access
to critical corporate applications and services. 
The most attractive feature
of the Tarantella system is its resumable applications. 
It allows users to log out of their webtops while keeping their applications running on servers. As 
a result, users can initiate a long-duration calculation, log out of 
the Tarantella system, and then find their results after reaching their 
destinations \cite{tarantella2005}. 
Our approach, on the other hand, addresses the problem associated with file downloading in 
heterogeneous wireless environments.
It focuses on resumable file downloading and network environment awareness, 
rather than on resumable lengthy computation.   

Wireless Application Protocol (WAP) aims to provide Internet content to mobile devices,
pagers and other wireless terminals over low-bandwidth wireless networks \cite{WAP2000}. 
The WAP stack is divided into five hierarchy levels: (1) wireless application environment (WAE); (2)
wireless session protocol (WSP); (3)  
wireless transaction protocol (WTP); (4) Wireless Transport Layer Security (WTLS)
and (5) wireless datagram protocol (WDP).
WAP places its premium on the efficiency of wireless data transfer. But this 
efficiency comes at a price. In order to leverage the popularity of Web servers, 
WAP needs to first translate the HTML documents to the corresponding WML (wireless markup language) 
documents.
Then it needs to convert the WML content into the binary format 
to further 
reduce the transmission size \cite{WAP2000}. Such protocol transformation
requires an additional processing step that considerably complicates the end-to-end 
information flow. 
The need for wireless efficiency of WAP is in fact gradually obviated 
by advances in wireless technologies such as 3G and IEEE 802.11b.  

Our prototype system addresses different aspects of wireless data transfer than 
the Tarantella system and the WAP protocol. 
The difficulty of vehicular Internet access is reflected in 
the complexity of network heterogeneity, and the autonomy of service providers 
and wireless carriers.  
In addition, transient network failures due to the presence of 
blind coverage spots and vehicle speed should be one of the primary concerns in 
vehicular network environments. 

To investigate the effect of vehicle speed on network connectivity and system throughput, 
we conducted a real-life experimental study. 
We used an ICMP-based probe with 
packet size of $64$ bytes to 
poll {\it www.yahoo.com} continuously 
using the Verizon CDPD and CDMA {\it 1xRTT} networks respectively. 
We then measured 
the round-trip time (RTT) of each probe. 
Figs(\ref{fig:figure0000})-(\ref{fig:cmdamobile})
present the measurement results obtained from 
a $45$-minute drive time during rush hour (9am-11am).    
Figs(\ref{fig:fixcdpd})-(\ref{fig:fixcdma}) represent 
the results obtained from a
fixed-location test 
(a window office around 11:3am). Notice that 
in Figs(\ref{fig:figure0000})-(\ref{fig:fixcdma}), 
transient network disconnects were 
represented by zero round trip time for a better visualization. 

\begin{figure}[htb]
\centerline{\psfig{file=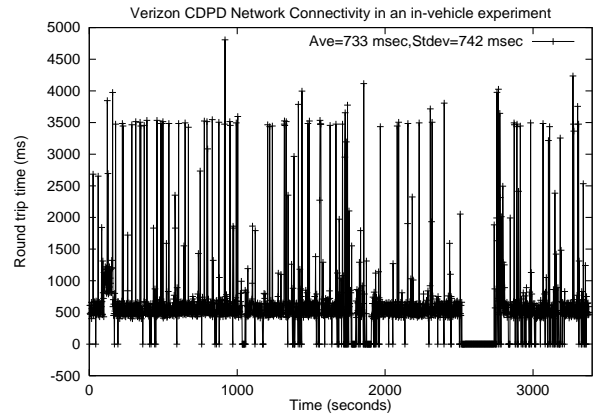,height=2.2in,width=3.2in}}
\caption{In-vehicle Verizon CDPD connectivity}
\label{fig:figure0000}
\end{figure}

\begin{figure}[t]
\centerline{\psfig{file=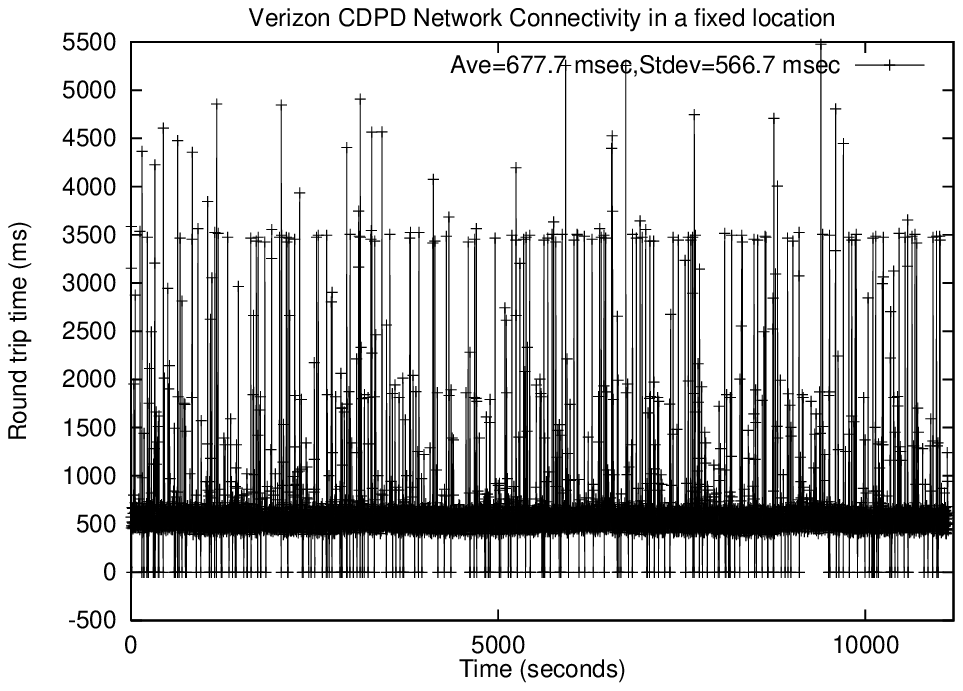,height=2.2in,width=3.5in}}
\caption{Fixed-location Verizon CDPD connectivity}
\label{fig:fixcdpd}
\end{figure}

\begin{figure}[htb]
\centering 
\centerline{\psfig{file=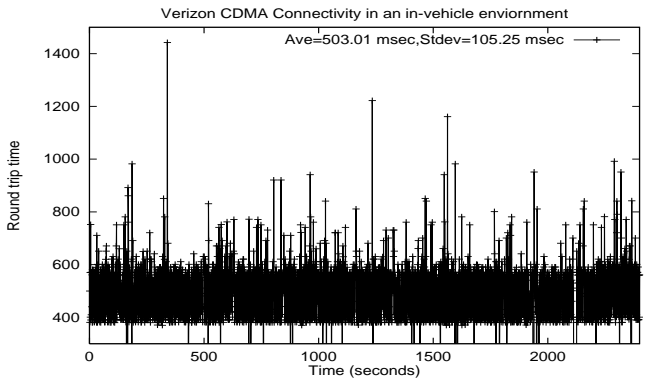,height=2in,width=3.5in}}
\caption{In-vehicle Verizon CDMA connectivity}
\label{fig:cmdamobile}
\end{figure}

\begin{figure}[htb] 
\centerline{\psfig{file=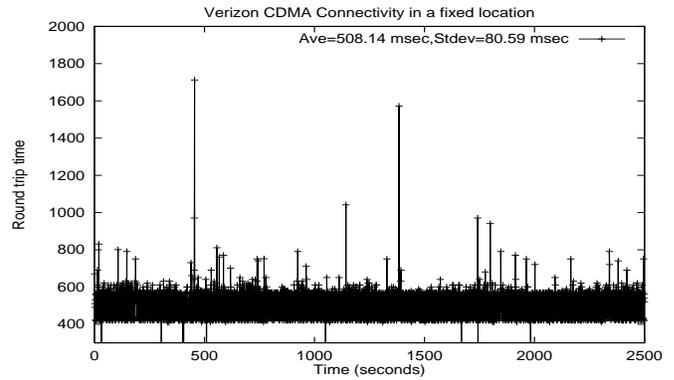,height=2in,width=3.5in}}
\caption{Fixed-location Verizon CDMA connectivity}
\label{fig:fixcdma}
\end{figure}

This comparison study showed 
that among the total $1689$ samples taken, 
$203$ packets were dropped in 
the in-vehicle CDPD connectivity experiment. This
accounts for a $12\%$ packet drop compared to the 
$194/5567=3\%$ packet drop rate in the fixed location.
Also among the total $2401$ samples taken, 
$58$ packets were dropped in the in-vehicle CDMA connectivity experiment, 
accounting for a $2.4\%$ packet drop
compared to the $9/2500=0.3\%$ packet drop rate in the fixed location.
We also observed that among the total $1689$ probes in the in-vehicle CDPD connectivity
experiment, $97 (5.74\%)$ probes were responded with 
no network connectivity due to the existence of blind coverage spots during driving.
By removing dropped packets and packets responded with 
no network connection,  we 
used the mean and standard deviation to  
quantify the delay variability in Table~\ref{tab:cdmadata}.

\begin{table}[htbp]
\begin{center}
\caption{Connectivity in fixed-location and in-vehicle environments}
\label{tab:cdmadata}
\begin{tabular}{||l|l|l|l|l||} \hline \hline
& \multicolumn{2}{c} {\em CDPD} & \multicolumn{2}{|c||} {\em CDMA} \\  \hline \hline
& mean &  stdev  & mean & stdev \\ \hline \hline
in-vehicle &  $733ms$ & $742ms$ & $503ms$  & $105ms$ \\ \hline
fix-location  & $677ms$ & $566ms$ & $508ms$ & $81ms$  \\ \hline \hline
\end{tabular}
\end{center}
\end{table}

The real-life experimental results in Table~\ref{tab:cdmadata} 
suggested that vehicle speed indeed introduces an additional 
delay variability (increased standard deviation of RTT in 
both CDPD and CDMA tests), and thus could be considered as a 
non-negligible cause of connectivity unreliability and 
performance deterioration.  
This research work is motivated by the fact that vehicular Internet access
is an important part of telematics applications, especially for various rear-seat applications. 
We identify several important but overlooked factors that negatively affect 
vehicular Internet access,
such as network hysteresis and frequently encountered network failures, 
and present a framework in an effort to mitigate these problems. 

\section{Architecture Description}
In this section,
we focus on the design and implementation of a prototype system to 
support an efficient HTTP failure recovery, network awareness and network 
adaptation. 
It is worth noting that our prototype system is built on top of 
a Microsoft Window 2000 server and XP. 
As a service subscriber to the Verizon CDMA data service, 
we do not have control over the Verizon CDMA network as well. 
   
It is well-known that Web servers and Web browsers are 
the fundamental architectural building blocks in the World Wide Web.
Web servers are intrinsically 
stateless and each request is processed without any knowledge of 
previous requests.  Any network failure will disrupt 
ongoing HTTP sessions, thus requiring 
the user to manually reestablish a connection to the same server. 
This HTTP failure recovery mechanism works well in the wired network
but it does not sit well with the wireless 
vehicular environments because 
of the 
lack of two essential features:
\begin{itemize}
\item automatic and efficient HTTP failure recovery without human intervention,
\item network awareness with ability to adapt to changing network conditions. \\
\end{itemize}
An automatic HTTP failure recovery means that the recovery is invoked as 
an ancillary event in response to a network failure, without any manual intervention on 
the part of the user.  This feature, however, is not supported by 
Web browsers.  
Furthermore, the real challenge is to adequately address the problems related to
vehicular environments
while at the same time keeping the existing Web server and 
Web client intact. 

To meet these requirements, 
we propose a multi-tiered architecture as shown in Fig(\ref{fig:figureflow}).
This architecture consists of a client-side proxy and an information gateway (IGW hereafter), 
sitting 
between the Web browser and the Web server.
The addition of the client-side proxy and the IGW is to provide a  
shield that supports automatic and transparent HTTP failure recovery while keeping
existing Web server and Web browser intact. 
The entire prototype system  
is structured  into 
the hierarchy functional layers.
The client-side proxy subsystem is implemented through
the layering of five types of technologies, and the IGW subsystem is implemented 
through the layering of four types of technologies.  
As a means of walling off complexity,
the layer classification focuses
on structural extensibility.
As a result, a new layer could be added flexibly without
needing to rewrite a significant chunk of infrastructure to
support the requirement change.  

\begin{figure}[htb]
\centerline{\psfig{file=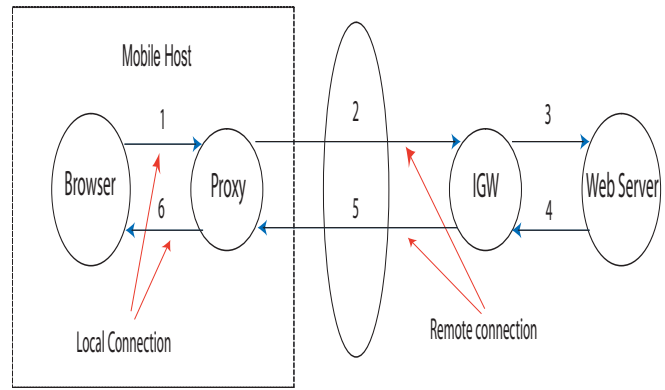,height=2in,width=3.4in}}
\caption{Processing Flow}
\label{fig:figureflow}
\end{figure}

A detailed breakdown of the layering structure is shown
in Fig(\ref{fig:figure030}). The top layer at MH is 
HTTP proxy
that provides a transparent access 
interface to the Web browser
by concealing complicated HTTP failure recovery and network awareness functionality.
Its primary task is to intercept HTTP requests from 
the Web client and to split connectivity into HTTP local and remote connections.
The HTTP session layer is to keep
track of each ongoing HTTP session (byte-count and time stamp of each ongoing session) 
and to trap various system events generated by the 
network sensing layer.
Upon the receipt of event notification such as network failure, 
the HTTP session layer 
can automatically 
restore the affected HTTP sessions by exchanging session information
with its counterpart at the IGW.
The network sensing layer can
capture changing network conditions and inform the HTTP session layer of such changes 
in a timely fashion.
In addition, it is able to dynamically select the best network interface in the 
presence of multiple networks, providing network awareness and adaptation. 

\begin{figure}[htb] 
	\centerline{\psfig{file=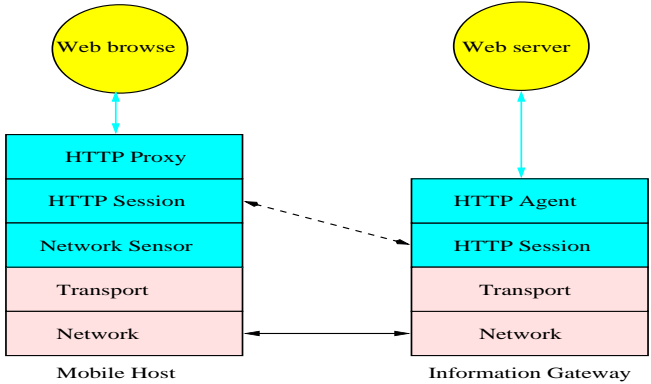,height=2in,width=3.4in}}
\caption{Logical Layering Hierarchy}
\label{fig:figure030}
\end{figure}

The HTTP session layer at the IGW 
works in tandem with its counterpart at the MH,
providing the stateful management of ongoing HTTP session.
In the event of network failures,  
the HTTP session layer at the MH automatically 
resynchronizes with the IGW with the stateful information 
about affected ongoing session without human intervention. 
This allows the HTTP session layer at 
IGW to pick up the data stream at point of interruption.
Finally, the HTTP agent layer at the IGW 
is to directly interface with Web servers.  
  
It is worth noting that FTP protocol with restart option and HTTP restart 
in HTTP 1.1 specification could be used to support packet-level failure recovery. 
Rather than focusing exclusively
on failure recovery protocols, our approach is based on a multi-tier  system architecture,  
focusing not only on automatic packet-level HTTP failure recovery, but also 
on network awareness and adaptation.
Our prototype system is widely applicable,
irrespective of the protocols being used.  
In the following sections, we will go through each layers of the hierarchy structure of each subsystem in detail.
\subsection{Network Sensing Layer \& HTTP Session Layer}
The network sensing layer at the MH
provides the network-aware capability, thus serving as 
a means to inform the HTTP session layer of 
changes in network conditions (see 
Fig(\ref{fig:figure09}) for details). 

The ability to rapidly discern changes in network conditions
and to pinpoint the root cause of such changes is 
an essential network-aware feature. It also 
plays a crucial role in automatic failure recovery.
To this end,  we consider 
two application-layer mechanisms as the core of network 
sensing layer: (1) event-driven scheme and (2) polling-based scheme.
These two complementary schemes are used in parallel in the prototype system, 
serving distinct roles in detecting 
various types of network failures. 

The event-driven scheme is used to 
capture various network events taking place 
during the course of HTTP sessions. 
From an application-layer perspective, 
ongoing HTTP sessions are fundamentally
{\it socket}-related. As a result,
socket-related exceptions are raised 
when network failures (events)  occur during HTTP sessions.
The idea underlying the event-driven scheme 
is to take the advantage of the {\it socket}-related exception handling
to trap various network failures (events) during the course of 
HTTP sessions and to 
identify the root cause of such (events) failures.

The polling-based scheme is used to gather 
the overall network conditions 
via periodic polling of  MH's network interfaces. 
This scheme is particularly useful for detecting 
changes in network environment when the MH moves into the coverage 
area of a new WLAN network in the absence of  active HTTP sessions.
In our implementation, an asynchronous thread-based {\it poller} 
is used to periodically 
retrieve network interface information on the MH,
using the {\it WSAIoctl} function  with  
{\small SIO\_GET\_INTERFACE\_LIST} option in Microsoft Platform SDK. 
The presence or disappearance of a wireless  network 
is detected through periodic polling  
by comparing the current status with the previous one. 
A notification to the HTTP session layer 
will be generated to report such a change. 

\lstset{basicstyle=\small,stringstyle=\ttfamily,language=C++}
\begin{lstlisting}[float,frame=tb,caption={Event Capturing},label=pseudocode]
try {
  SocketConnection
  while ( EOF == false)  
  { 
     WriteSocket
     nRead=ReadSocket
     bytecount = bytecount+nRead;
     if a higher bandwidth network found
         return PREEMPTIVE_EVENT;
   }
   return;  
  } 
  catch(SocketException) 
  { 	
    return GetExceptionCode;	
  }
 \end{lstlisting}

\lstset{basicstyle=\small,language=C++}
\begin{lstlisting}[float,frame=tb,caption={Exception Handling},label=exception]
switch (nRet)
{
  case 0:   
    request = GetRequest();
    break;                	
  case PREEMPTIVE_EVENT:
    Preemptive();
    break;  
  case WSAEHOSTDOWN:  
  case WSAECONNABORTED:
  case WSAECONNRESET:
  case WSAENETDOWN:
  case WSAENETUNREACH:
  case WSAENETRESET:
  case WSATRY_AGAIN:
  case WSANO_RECOVERY:
  case WSAEADDRNOTAVAIL: 
    Handoff();
    break;
  default;
    break;
}
\end{lstlisting}

The pseudocode in Listing~\ref{pseudocode} 
illustrates our implementation of the event-driven scheme
to capture random network events during an HTTP session.
The code within the {\it try block} section corresponds to 
an ongoing HTTP session being executed. 
A network event taking place during the 
course of a HTTP session will trigger 
an exception in the {\it try block} section.  
The code within the 
{\it catch block} (exception handler) 
is thus invoked. The root 
cause of the network 
event is then identified by 
examining the corresponding exception error code.
In addition, we 
define a message code {\it PREEMPTIVE\_EVENT}
to represent the presence of a high bandwidth network during 
an HTTP session, 
which 
is detected by a thread-based {\it poller} via periodic polling.
The presence of such events results in a preemptive handoff (switching to 
a relatively higher bandwidth network) to better utilize the available bandwidth resource. 
The pseudocode  in Listing~\ref{exception}
describes how to identify the cause of network events (failures)
and how to invoke a proper function based on the nature of network event. 
A variety of error codes representing various transient network failures 
during an HTTP sessions is identified and provided in the above code snippet. 
\begin{figure}[htb]
\centerline{\psfig{file=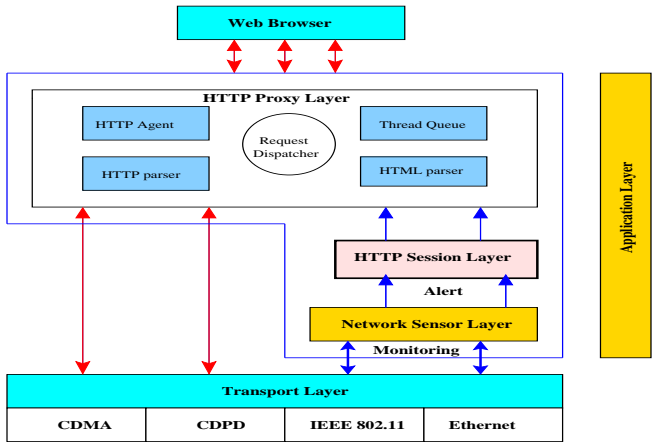,height=2.3in,width=3.4in}}
\caption{System architecture of mobile host}
\label{fig:figure09}
\end{figure}

Detection delay
is an important metric for measuring
the responsiveness of a MH to change in the network environment. 
An increased polling frequency generally results 
in a decreased detection delay, as shown in Fig(\ref{fig:delay}).
On the other hand, 
an excessive polling takes up run-time kernel resources \cite{YiBingLin2002}.
Thus it is important to select an appropriate  polling frequency 
to balance system responsiveness and run-time resources.
The following theorem establishes a connection between 
detection delay,  polling frequency and change rate of network environment,
by assuming that change in the network condition follows 
the Poisson process with intensity $\lambda$.   \\
\begin{theorem} \label{theorem:delay}
Let $T>0$ be a polling interval, and change in 
network conditions be a Poisson process with intensity $\lambda$, 
then the long-run mean average detection delay, $E(D)$,
is bounded by
\begin{equation} 
E(D) < \frac{(T\lambda)^2-2T\lambda +2-
2\exp(-T\lambda)}{2T(\lambda)^2}	,
\end{equation}
where $E(.)$ is the expectation function.  \hfill $\blacktriangle$ \\
\end{theorem}

Proof of Theorem is given in Appendix. 
Notice that polling interval is a tunable system parameter.
Fig(\ref{fig:delay}) presents the dependency of  
detection delay on polling frequency and network environment change rate.
Assuming that $T=1/\lambda = 10$ seconds, 
that is, the polling interval is the same as
the mean interarrival time of network condition of $10$ seconds, 
then the long-run mean average detection delay based on Theorem~\ref{theorem:delay}
is less than $10*(0.5-\exp(-1))= 1.34$ seconds. 
In our implementation, we set the polling time as $10$ seconds,
by taking into account  the responsiveness and the resource usage.

The primary function of HTTP session layer 
is to keep track of HTTP sessions, including the 
byte-count of each ongoing sessions as illustrated in 
the pseudocode  in Listing~\ref{pseudocode}.
The byte-count information plays a crucial role for packet-level 
HTTP failure recovery.    
\begin{figure}[htb]
\centerline{\psfig{file=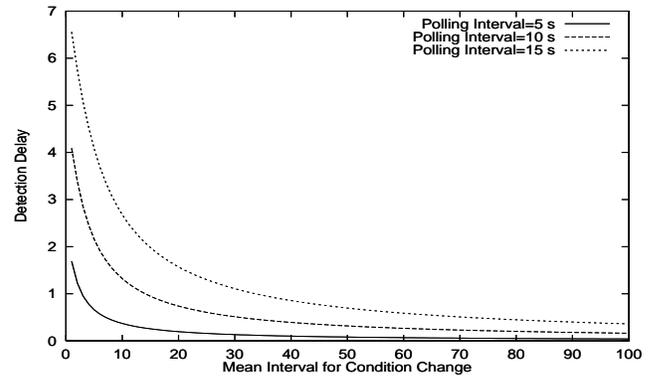,height=2in,width=3.4in}}
\caption{Detection Delay vs. Network Change Rate}
\label{fig:delay}
\end{figure}

\subsection{HTTP Proxy Layer}
The function of the HTTP proxy layer is 
to provide Web access transparency as well as to add a new functional
layer that supports network adaptation and packet-level failure recovery,
without the need to modify existing Web server and Web browser. 

The proxy layer
splits each HTTP request initiated from the Web browser 
into two separate HTTP connections: the local one to the Web client and 
the remote one to the IGW, as shown in Fig(\ref{fig:figureflow}).
Upon receipt of an HTTP request from the Web browser, 
the client-side proxy  transparently redirects the request to the IGW, which
in turn forwards the request to 
an origin Web server.
The local connection 
between the proxy and the Web browser
is impervious to network failure,
whereas the remote connection
between the client-side proxy and the IGW is vulnerable to 
network failure. 
Such a splitting connectivity
provides a clear separation of  local and remote HTTP sessions,
allowing us to gracefully handle network failures 
while at the same time maintaining user-perceived 
HTTP continuity.
Additionally, 
the packet-level HTTP failure recovery is 
implemented into the HTTP proxy layer in which  
fragmented data is properly pieced together to avoid restarting the HTTP session from scratch 
in the presence of network failures. Thus it ensures
the efficiency of HTTP failure recovery without the need to  modify
existing Web client and server.

\begin{figure}[htb]
	\centerline{\psfig{file=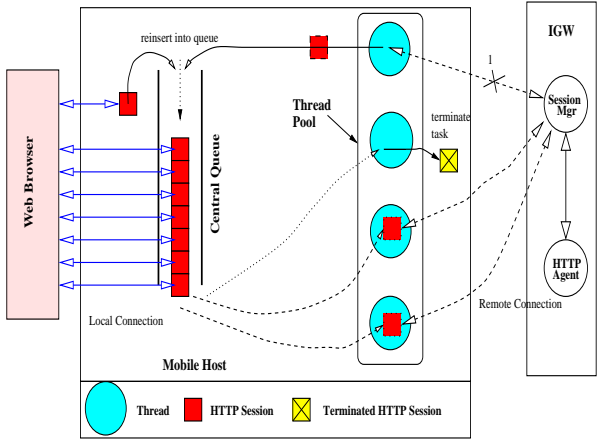,height=2.3in,width=3.1in}}
\caption{Thread Pool and Task Queue}
\label{fig:figure0001}
  \end{figure}

An instance of the HTTP session class, which 
is derived directly from Microsoft foundation class {\it CObject},  
is created for per HTTP request 
and is terminated (released) when the HTTP session has completed.  
The class object contains the two socket objects for local and remote connections.
Since it is connected to the Web browser via loopback IP address $127.0.0.1$, the local 
connection is 
persistent across the life time of the object. 
The remote connection, however, is independent of the life time of the thread object;
it could be terminated and recreated several times across the 
duration of the class object, depending on network conditions. 

We implemented a thread pool architecture 
\cite{Ling2000,Lewis1996} to efficiently
manage worker threads.  
The idea behind the thread pool is
to reuse worker threads. 
Each worker thread in the pool thread can be 
recycled, thereby improving the system performance by
avoiding repeated and costly thread creation. 
Upon startup, a fixed number of worker threads are created waiting for
tasks. A thread object in the queue is extracted by an idle worker thread 
in the pool in the FIFO fashion.
It can be seen from Fig(\ref{fig:figure0001}) that
a newly created thread object, which corresponds 
to an HTTP request initiated by the Web client,  
is immediately put at the end of a queue 
waiting for processing. 
As the worker threads finish with old tasks 
and become available, 
a thread object is extracted from the queue. 
When a network failure (handoff) occurs during HTTP sessions, 
affected thread objects
are placed at the end of the queue again waiting for the network to recover
and will be reprocessed
by an idle worker thread in the thread pool.
The lines 1 and 2 in Fig(\ref{fig:figure0001})
represent the reprocessing flow. 
The size of the thread pool (number of worker threads) 
in our implementation is set to four
for a balanced 
threading concurrency and context switching overhead  
\cite{Ling2000,Lewis1996}.

\subsection{Implementation of the IGW}
At the heart of the IGW architecture, it implements 
a task dispatcher component as illustrated in Fig(\ref{fig:igwarchi}), which 
is based on  an ISAPI extension directly inherited from Microsoft 
{\it CHttpServer} class.
  \begin{figure}[htb]
	\centerline{\psfig{file=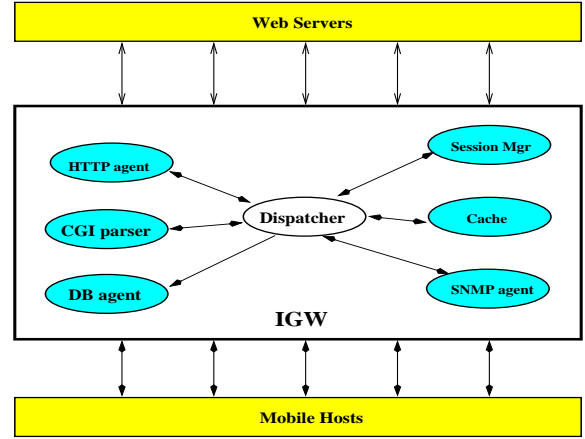,height=2.3in,width=3in}}
\caption{Processing Flow}
\label{fig:igwarchi}
\end{figure}

The task dispatcher, which is implemented on top of the Microsoft {\it CHttpServer} class, 
handles each incoming HTTP request, 
parses request message, and decides which  
action to take (which agent to invoke). 
An instance of {\it CGI} object is created upon the receipt of each
HTTP request, with its pointer to an 
{\it EXTENSION\_CONTROL\_BLOCK} structure defined by 
Microsoft SDK.  The  CGI 
object implements a parser for retrieving 
various information embedded in the HTTP request header. The HTTP agent 
is then used to retrieve HTTP files from origin Web servers on behalf of 
MHs. In the case of HTTP failure recovery,  the offset information
retrieved by the CGI parser
is passed to  the HTTP session manager for repositioning the begin offset for 
file downloading.  The pseudocode  is given in 
Listing~\ref{Exserver}. 

\lstset{basicstyle=\small,language=C++}
\begin{lstlisting}[float,frame=tb,caption={GetExServerConnection},label=Exserver]
 GetExServerConnection() 
 {
   parse HTTP request
   extract session offset 
   connect to a Web server;
   forward request to Web server
   retrieve content at begin offset 
   while (EOF == FALSE) {
      send content to  MH 
      read content from Web server; 
  }
  send content to the MH 
};
\end{lstlisting}
The following example is given to illustrate how this scheme works.
Assume that the MH sends a request to retrieve a 
document with the URL
http://www.cnn.com/draft.ppt. 
The HTTP request header is listed in Table~\ref{tab:request}:

\begin{table}[htb] 
\begin{center}
\caption{Origin HTTP Request Header}
\label{tab:request}
\begin{tabular}{|l|} \hline 
GET http://www.cnn.com/draft.ppt HTTP 1.0 $\backslash$r$\backslash$n  \\
$\cdots$  \\
$\backslash$r$\backslash$n$\backslash$r$\backslash$n \\ \hline
\end{tabular}
\end{center}
\end{table}

Upon receiving the request initiated by the Web browser, 
the client-side proxy makes a transparent redirection 
to an IGW with IP address $205.132.6.11$, 
by adding session information into the HTTP request header,  as showed in table~\ref{tab:modified}.

\begin{table}[htb] 
\begin{center}
\caption{Modified HTTP Request Header}
\label{tab:modified}
\begin{tabular}{|l|} \hline 
GET http://205.132.6.11/scripts/dis.dll?url=http://www.cnn.com/\\
draft.ppt HTTP 1.0 $\backslash$r$\backslash$n  \\
User-Agent: Proxy/2.0 $\backslash$r$\backslash$n \\
Session-Offset: $0$ $\backslash$r$\backslash$n \\
$\cdots$ \\
 $\backslash$r$\backslash$n$\backslash$r$\backslash$n \\ \hline 
\end{tabular}
\end{center}
\end{table}

Notice that {\it dis.dll} is the binary executable of the IGW implementation. 
The modified URL means that the source of content is located at 
{\it http://www.cnn.com/draft.ppt}; and that
the document should be downloaded from scratch (Session-OffSet: 0). 
Upon receiving the request,
the IGW  extracts the source URL contained in the modified HTTP 
request header and sends the request to the origin Web server (http://www.cnn.com/draft.ppt). 
The response from the server will be relayed   
to the client-side proxy at the MH
(see Fig(\ref{fig:figureflow}) for details). 

Assume that a network handoff or preemptive handoff occurs 
during the course of file downloading, which 
disrupts the ongoing HTTP session. 
The number of bytes received by the MH so far is assumed to be $203,223$ bytes. 
The byte-count information is added into an HTTP 
request header (offset=$203,223$), as showed in Table~{\ref{tab:networkfailure}).
When the client-side proxy automatically reestablishes 
a remote connection to the IGW, which in turn  
will act accordingly by fetching data from the $203,223$ bytes
from the beginning of the file (draft.ppt).  
Data packets received by the client proxy at the MH will be properly 
pieced together to form a complete powerpoint document.  

\begin{table}[htb] 
\begin{center}
\caption{Modified HTTP Request Header After Network Failure}
\label{tab:networkfailure}
\begin{tabular}{|l|} \hline 
GET http://205.132.6.11/scripts/dis.dll?url=http://www.cnn.com/\\
draft.ppt HTTP 1.0 $\backslash$r$\backslash$n  \\
User-Agent: Proxy/2.0 $\backslash$r$\backslash$n \\
Session-Offset: $203223$ $\backslash$r$\backslash$n \\
$\cdots$ \\
 $\backslash$r$\backslash$n$\backslash$r$\backslash$n \\ \hline 
\end{tabular}
\end{center}
\end{table}

The process of 
failure recovery is performed between the client proxy and the IGW 
in an automatic and transparent fashion without human interference.  

\section{Experimental Studies}
In this section, 
we first describe our testbed which closely resembles 
a real heterogeneous wireless environment,
then present the experimental results of the prototype system.

The entire implementation of client-side proxy and IGW subsystems 
consists of roughly $20,000$ and $8,000$ 
lines of Visual C++ source code, respectively. 
On the MH, approximately $10,000$ lines of code are GUI-related, 
The remaining lines of code are written specifically for
thread pool, HTTP parser, network awareness, and HTTP session management.
On the IGW, the code primarily deals with HTTP agent, CGI parser, database agent,  
and HTTP session management. 

The testbed consisted of two Toshiba Tecra laptops (MHs) running 
Microsoft XP operating system and one Dell OptiPlex desktop (IGW) 
running Microsoft $2000$ server with $192$ MB main memory and 
{\it x}86 family 
6 Model CPU. 
Each Toshiba Tecra laptop has
$1$ GB main memory with an Intel
Pentium III $1.2$ GHz Mobile CPU and 
the built-in WLAN card with maximum rate of $11$ Mbps. They also have
a switch below the keyboard on the front, which can be used  to 
manually  turn on or off the built-in WLAN card. 
In addition, we installed Verizon CDMA {\it 1xRTT} 
driver with the Sierra wireless AirCard 550 
with maximum rate of $144$ Kpbs.
The routing cost metrics for the Ethernet, WLAN and 
CDMA interfaces are configured as  
$1$, $2$, and $5$, respectively, so that the network interface with 
lower cost metric will be selected 
in the presence of multiple network interfaces.    

A WavePoint II access point outfitted
with $10$ Dbi Omni antenna
was used as an IEEE 802.11b base station to serve as a bridge between the WLAN and 
the wired network.
It was attached to the side of a  window facing east 
in a window office (in Applied research building in Morristown, NJ),  
offering a limited wireless outdoor coverage through the window. 
Under this testbed configuration, 
the IGW was accessible either from the Verizon CDMA network over $17$ hops 
or from the WLAN over $2$ hops. 
The entire driveway was under the coverage of 
the Verizon CDMA network, but only part of the driveway was
under the coverage of the WLAN (see Fig(\ref{fig:testbed}) for details). 

To evaluate the prototype system in a vehicular 
network environment, we conducted the experimental study 
focusing on the following three aspects: 
(1) network environment awareness, 
(2) transparent packet-layer HTTP failure recovery 
in the presence of handoffs,
and (3) preemptive handoff by dynamically selecting the best network interface in the overlapping coverage area
of the CDMA and the WLAN. Notice 
that the MHs in our testbed were always in the coverage 
of the Verizon CDMA {\it 1xRTT} network, 
whereas the WLAN coverage relied purely 
on the location with respect to the WLAN's access point. 

\begin{figure}[htb]
\centerline{\psfig{file=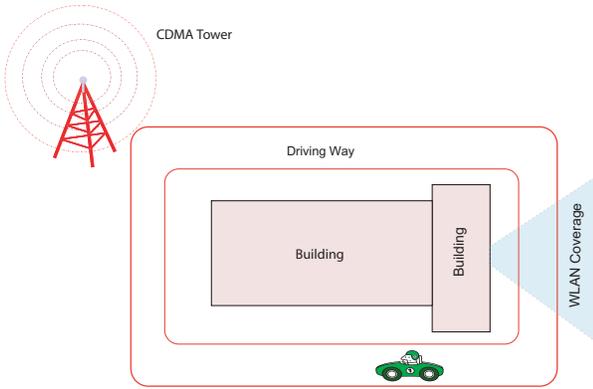,height=2in,width=3.1in}}
\caption{Testbed Environment}
\label{fig:testbed}
\end{figure}

To establish a baseline comparison, we placed  
two MHs with/without software installation
in the same vehicle.
The vehicle's initial position was under the 
coverage area of the WLAN. We initiated a long-lived HTTP session 
on both MHs at the same 
time (downloading a $6$M powerpoint file) and then drove the vehicle circling 
the building with speed approximately $3\!\sim\!7$ mph,
thereby creating an intermittent WLAN connectivity during file downloading.
The side-by-side comparison demonstrates that
the MH having this prototype system 
can resynchronize with the IGW in the presence of handoffs and  
can dynamically select the best available network interface 
in the overlapping coverage area,  
while at the same time maintaining an 
user-perceived HTTP session continuity.  
In contrast, a MH without this prototype system was unable to 
maintain HTTP session when handoffs occurred. 

\subsection{Performance Comparison with and without preemptive handoff}
It is not trivial to quantify the performance gain of the preemptive 
handoff in an in-vehicle environment, 
because the vehicle's motion, in effect, is not repeatable.   
To simulate the intermittent WLAN connectivity  
in a repeatable fashion, we conducted
the experimental study in a 
window office with 
a good reception of the Verizon CDMA network. 

A series of experiments under different network settings 
was designed to investigate the impact of network 
hysteresis on system throughput.
First, we tested a system without network-aware capability,
Secondly, we tested the prototype system using   
the CDMA/WLAN 
and CDMA/Ethernet heterogeneous networks, respectively.

In the experimental study, 
one MH initially had only 
the Verizon CDMA network connectivity.
A long-lived HTTP session was initiated
to download a $500K$ powerpoint document over 
the Verizon CDMA {\it 1xRTT} network. After $30$ seconds,
a high-bandwidth network was activated by either manually plugging 
the Ethernet cable into the MH
or turning on the MH's WLAN card,
in an effort to emulate that the MH had suddenly 
moved into an overlapping coverage area of the WLAN.  
Our approach to control timing of network availability 
is similar to 
the experimental studies presented in \cite{Baker1996,stemm98vertical}.
We measured and compared the elapsed time needed to retrieve the file
with and without the preemptive handoff. 

\begin{table}[htbp]
\begin{center}
\caption{Performance comparison in heterogeneous network environments}
\label{tab:preemptive}
\begin{tabular}{||l|l|l|l|l|l||}  \hline \hline
\multicolumn{6}{||c||} {\bf Overall Time} \\ \hline
\multicolumn{2}{||c} {\em without preemptive} & 
\multicolumn{2}{|c} {\em preemptive} &
\multicolumn{2}{|c||} {\em preemptive}  \\ 
\multicolumn{2}{||c} {\em } & 
\multicolumn{2}{|c} {\em CDMA-WLAN} &
\multicolumn{2}{|c||} {\em CDMA-Ethernet} \\ \hline 
mean &  stdev  & mean & stdev  & mean & stdev\\ \hline 
$521.4s$ & $59.13s$ & $41.3s$  & $6.34s$ & $43.0s$ & $2.94s$\\ \hline \hline
\multicolumn{6}{||c||} {\bf discovery delay} \\ \hline
\multicolumn{2}{||c} {\em without preemptive} & 
\multicolumn{2}{|c} {\em preemptive} &
\multicolumn{2}{|c||} {\em preemptive}  \\ 
\multicolumn{2}{||c} {\em } & 
\multicolumn{2}{|c} {\em CDMA-WLAN} &
\multicolumn{2}{|c||} {\em CDMA-Ethernet} \\ \hline 
mean &  stdev  & mean & stdev  & mean & stdev\\ \hline \hline
{\em -} & {\em -} & $2.2s$  & $0.24s$ & $6.3s$ & $2.6s$\\ \hline \hline
\multicolumn{6}{l} {\em Notice: we used a fixed IP for WLAN and } \\
\multicolumn{6}{l}{\em a dynamically-assigned IP for Ethernet} 
\end{tabular}
\end{center}
\end{table}

Fig(\ref{fig:nopre}) presents
a snapshot of network utilization of CDMA and WLAN in the entire period of an HTTP session. 
It can be seen from Fig(\ref{fig:nopre}) that 
the MH initially had only CDMA network connectivity 
with the network utilization of $6\%$.  
Then a WLAN network was enabled at $30$ seconds 
after an HTTP session was initiated over the CDMA network. 
Without network-aware and preemptive capability, 
the MH continued to stay with the CDMA network
until the file had finished downloading
despite the presence of WLAN network 
during the HTTP session. As a result,
the network utilization of the WLAN was close to zero 
(Fig(\ref{fig:nopre}) for details).
In this case, it took up $521$ seconds on average 
to download a $500K$ powerpoint file over the CDMA network.
This is because the inherent network hysteresis prevents
the MH from taking advantage of the
available WLAN network during the session, 
once the HTTP session was initially established over the 
CDMA network.  

On the other hand, the MH enhanced with the 
preemptive handoff capability was able to sense as well as
to dynamically select a higher bandwidth network during the course of 
HTTP sessions, resulting in
a substantial saving in elapsed time of HTTP sessions.
Fig(\ref{fig:cmda-wlan}) and Fig(\ref{fig:cdma-ethernet})
show a spike in both WLAN, and Ethernet utilization 
rate after the preemptive 
handoff had taken place.   
Table~\ref{tab:preemptive} presents
the average and standard deviation of 
the overall down time and discovery delay over   
$10$ independent runs, with and without preemptive handoffs.  
There are two interesting points to note:
\begin{enumerate}
\item When both WLAN and Ethernet were
configured to activate at $30$ seconds after an HTTP 
session was initiated over 
the CDMA network, a preemptive handoff 
really took place after few seconds.
Such a delay could be
attributed to 
the amount of time required by OS to generate 
the connect event and to acquire an IP address 
dynamically assigned by a DHCP server;
\item  there existed an overlapping time
interval in which the MH continued to receive data 
packets from the CDMA network interface after 
it has actually closed a connection over the CDMA network and 
reestablished a connection over the WLAN. This portion of data 
was being labeled as 
useless traffic in Fig(\ref{fig:cdma-ethernet})
and Fig(\ref{fig:cmda-wlan}).
Such a phenomenon, being associated with the implementation of  
{\it TCP Time-Wait}, can be clearly explained in TCP protocol documentations
{\it RFC 793} and {\it RFC 1337}.
\end{enumerate}
\begin{figure}
  \begin{minipage}[htb]{3.5in}
	\centerline{\psfig{file=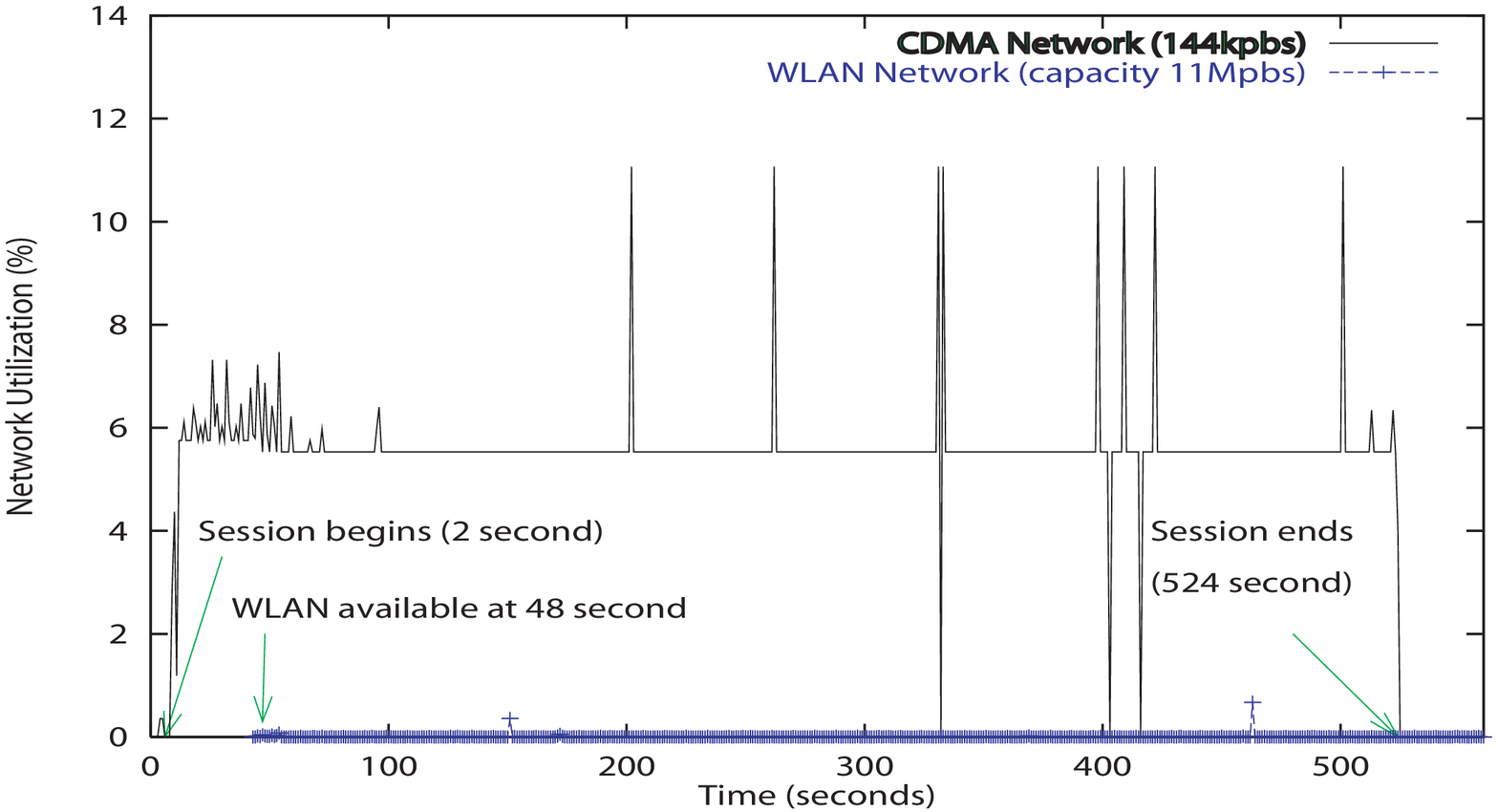,height=2in,width=3.5in}}
	\caption{Network Hysteresis (ignoring the presence of a higher bandwidth network during an HTTP session)}
         \label{fig:nopre}
  \end{minipage} 
\hspace{.04in}
  \begin{minipage}[htb]{3.5in}
	\centerline{\psfig{file=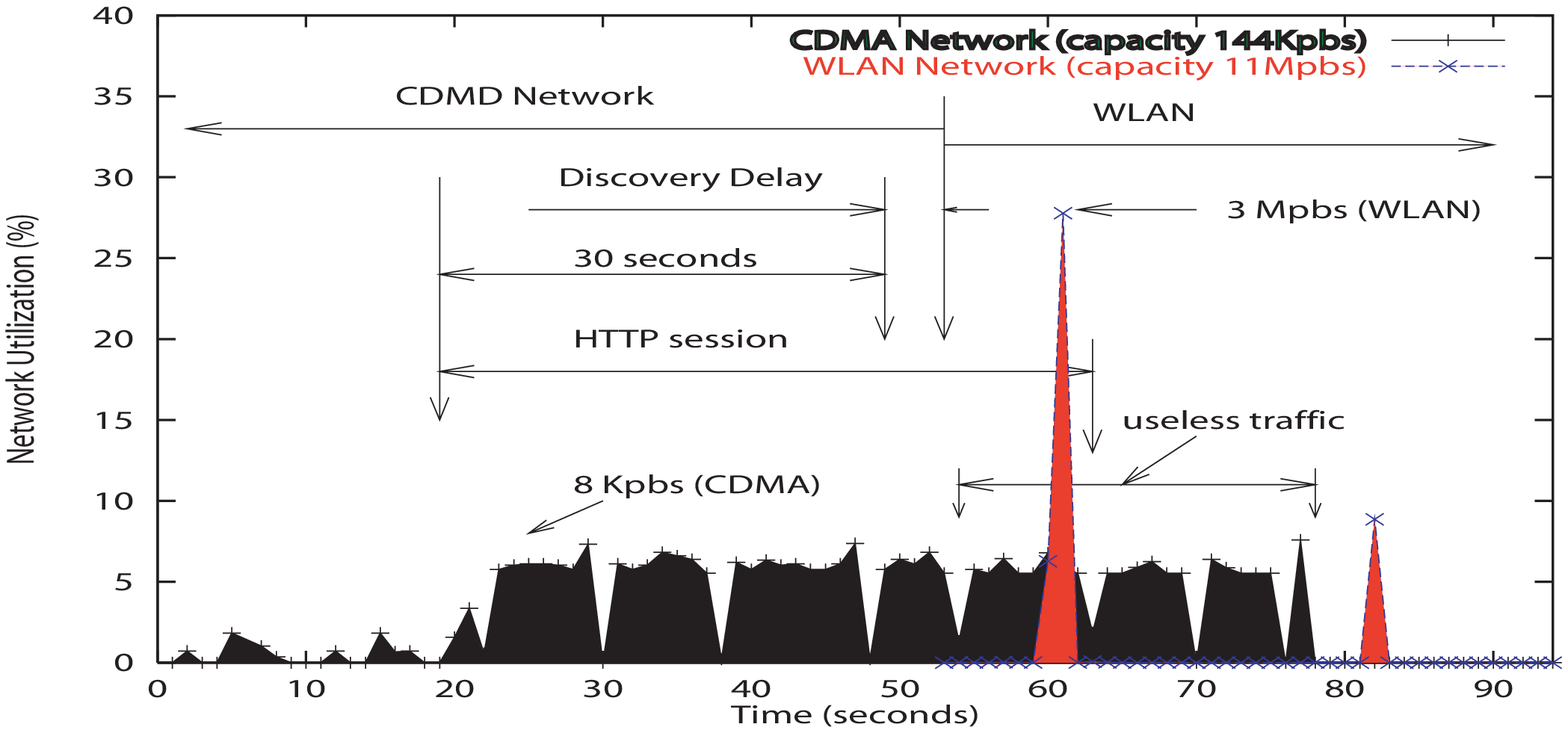,height=2in,width=3.5in}}
           \caption{Preemptive Handoff between CDMA and WLAN}
\label{fig:cmda-wlan}
  \end{minipage} 
\hspace{.04in}
  \begin{minipage}[htb]{3.5in}
	\centerline{\psfig{file=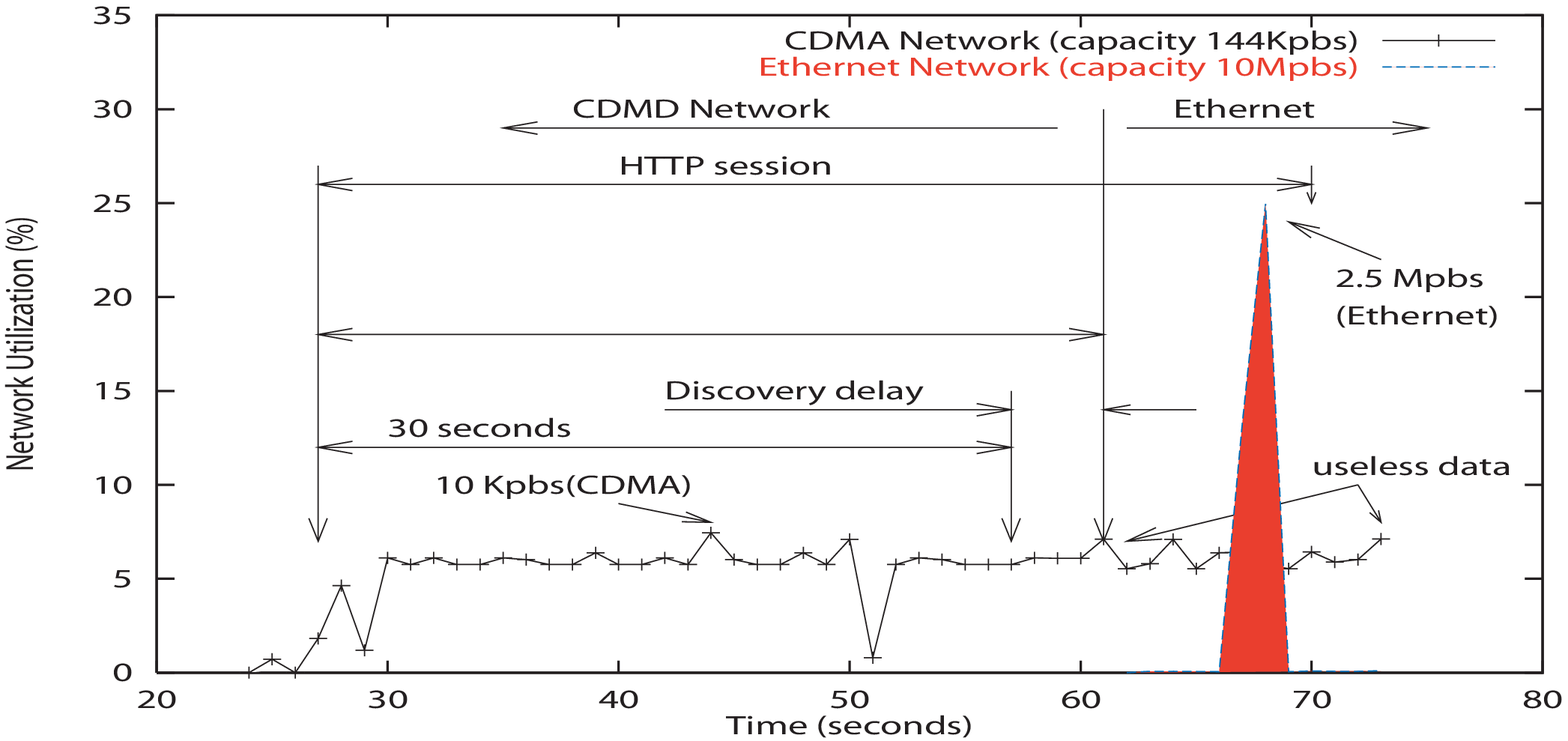,height=2in,width=3.5in}}
           \caption{Preemptive Handoff between CDMA and Ethernet}
\label{fig:cdma-ethernet}
  \end{minipage} 
\end{figure}

\begin{table} 
\begin{center}
\caption{diminishing benefit of preemptive handoff}
\label{tab:wlan-ethernet}
\begin{tabular}{||l|l|l|l|l|l||} \hline \hline
\multicolumn{2}{||c} {\em WLAN}&
\multicolumn{2}{||c} {\em Ethernet }&
\multicolumn{2}{||c||} {\em preemptive} \\
\multicolumn{2}{||c} {\em } &
\multicolumn{2}{||c} {\em } &
\multicolumn{2}{||c||} {\em WLAN-Ethernet} \\ \hline \hline
mean & stdev &mean &  stdev  & mean & stdev \\ \hline \hline
$41.3$ & $6.34s$ & $33.7s$ & $1.13s$ &
$44.8$ & $2.45$ \\ \hline \hline
\end{tabular}
\end{center}
\end{table}

The measurement results in Table~\ref{tab:preemptive} 
showed that in general the preemptive handoff significantly
improves the system throughput in an overlapping coverage area  
where two networks with high bandwidth differential are involved. 
However, the performance 
advantage of the preemptive handoff might be  
diminished or even become negative when the 
two involved networks have comparable bandwidth. 
For instance, we performed an experiment with
WLAN/Ethernet combination.  
As a baseline comparison, 
we first measured the elapsed time to download a $23$M powerpoint file
purely from the WLAN ($11$ Mbps) and 
the Ethernet ($10$ Mbps), respectively. 
To study the impact of preemptive handoff, 
one MH was initially set to have the WLAN connectivity. 
An HTTP session to download a $23$M powerpoint file 
was started with the WLAN for $15$ seconds,  
then a preemptive handoff from the WLAN to 
the Ethernet was triggered by manually plugging 
the Ethernet cable into the MH.
\begin{figure}[htb]
\centerline{\psfig{file=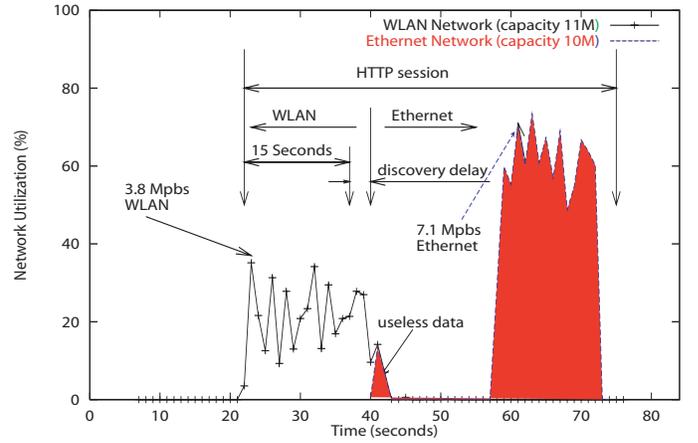,height=2.3in,width=3.5in}}
 \caption{Preemptive Handoff between WLAN and Ethernet}
    \label{fig:wlan-ethernet}
  \end{figure}
Under this setting, the negative effect of the
preemptive handoff was illustrated in
Fig(\ref{fig:wlan-ethernet}) 
and Table~\ref{tab:wlan-ethernet}. 
This effect was observed because that it normally took a few 
seconds of time delay to complete a preemptive handoff.
In general,        
the net benefit of the preemptive handoff can be formulated as  
\begin{align}\label{eq:eq1}
T^A_{est} > 
T^B_{est} + T_{handoff},
\end{align}
where the subscript $est$ denotes estimate remaining download time and 
the subscript $handoff$ denotes handoff time
and the superscripts $A$ and $B$ refer to networks A and B. 
Eq(\ref{eq:eq1}) means that a preemptive handoff 
from network A to network B makes
sense if the 
estimated remaining download time using network A 
should be greater than the estimated remaining download time using network B plus
handoff time.  In general, knowledge of the remaining size of an
ongoing HTTP session, 
together with the 
elapsed time for network switch, is an important factor 
that affects the efficiency of preemptive handoff.
The dynamic load of network has an effect on 
the end-to-end delay and actual network throughput. Hence,   
the real-time loads of involved networks are also an important 
factor for preemptive handoff, 
which adds another dimension of complexity. 

\subsection{Experiment for System Resilience}
We conducted another experiment to
test the resilience of the prototype 
system to random
transient network disconnects. 
To this end, in the experimental study we artificially create  
a random network failure in the midst of  HTTP sessions to see how 
the prototype system responds to it. 
Three cases are considered in our experiment:
\begin{enumerate}
\item we measured the time for 
downloading a $6$M file via the WLAN without network failure.
\item  a random network failure was 
produced by manually turning the AP's power off, then
turning the AP's power on,
\item  a random network failure 
was created by 
manually turning off the MH's WLAN card,
followed by turning on the MH's WLAN card.
\end{enumerate}
The first case served as a baseline for comparison study.  
We measured the download time of a $6$M file via either 
a dynamical IP address assigned by the DHCP server or a 
static IP address. 

\begin{table} [htb] 
\begin{center}
\caption{Impact of Transient Failures}
\label{tab:poweroff}
\begin{tabular}{||l|l|l|l|l|l||} \hline \hline
\multicolumn{6}{||c||}{\bf fixed IP address} \\ \hline \hline
\multicolumn{6}{||c||}{\bf overall download time} \\ \hline
\multicolumn{2}{||c} {\em No disruption}&
\multicolumn{2}{|c} {\em AP power off-on} & 
\multicolumn{2}{|c||} {\em WLAN off-on} \\  \hline   
mean & stdev &mean &  stdev  & mean & stdev \\ \hline 
$12.3$ & $3.16$ &
$101.9s$ & $3.93s$ 
& $43.5s$  & $3.69s$ \\ \hline  
\multicolumn{6}{||c||}{\bf network disconnect time} \\ \hline
mean & stdev &mean &  stdev  & mean & stdev \\ \hline
$-$ & $-$ &
$86.0s$ & $4.21s$ & $25.3s$  & $4.4s$ \\ \hline \hline \hline \hline
\multicolumn{6}{||c||}{\bf dynamically assigned IP address} \\ \hline 
\multicolumn{6}{||c||}{\bf overall download time} \\ \hline
\multicolumn{2}{||c} {\em No disruption}&
\multicolumn{2}{|c} {\em AP power off-on} & 
\multicolumn{2}{|c||} {\em WLAN off-on} \\  \hline 
mean & stdev &mean &  stdev  & mean & stdev \\ \hline
$12.3$ & $3.16$ &
$101.7s$ & $1.77s$ & $97.1s$  & $23.5s$ \\ \hline 
\multicolumn{6}{||c||}{\bf network disconnect time} \\ \hline
mean & stdev &mean &  stdev  & mean & stdev \\ \hline
$-$ & $-$ &
$84.7s$ & $3.86s$ & $74.7s$  & $17.7s$ \\ \hline \hline
\end{tabular}
\end{center}
\end{table}

The experimental study showed that the prototype system can survive 
long network disconnects triggered by manually turning off the MH's WLAN
card or the AP's power.   
Table~\ref{tab:poweroff} showed the mean and 
standard deviation of the 
elapsed time of HTTP session over 
$10$ independent runs.
It can be seen in Table~\ref{tab:poweroff} and Fig(\ref{fig:dynamic})
that even though 
network disconnect time lasted up to $86$ seconds, the 
prototype system was able to automatically resynchronize with the IGW
once network connectivity was resuscitated while maintaining an user-perceived 
HTTP session continuity.  

It can be seen from Fig(\ref{fig:dynamic})
that the MH took more than $10$ seconds 
to capture the event that the WLAN card  had been shut down manually, 
and only took roughly two seconds to capture the disconnect event 
that the WLAN card has
been manually turned on. This observation is in line 
with Microsoft design guidance \cite{Micrsoft2002} which states 
that the NIC need to wait $10$ seconds before 
generating any disconnect event 
and the connect event is generated at most $2$ seconds.
As an application-layer approach, the prototype system cannot control 
the OS kernel to improve the system responsiveness to such events. 
It, however, demonstrates a resilience to network disconnects
that last more than $10$ seconds \cite{Anne1999,Fikouras99,Fikouras1999}.
 
\begin{figure}[htb]
	\centerline{\psfig{file=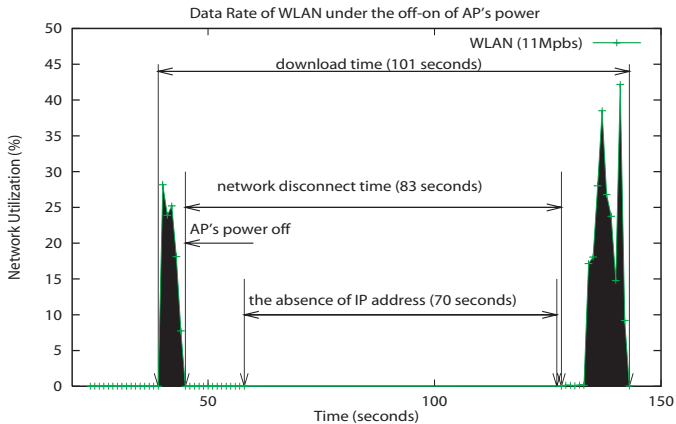,height=2.2in,width=3.5in}}
	\caption{HTTP Session under the off-on of AP's power}
\label{fig:dynamic}
\end{figure}

\subsection{Evaluation of Horizontal Handoff}
The goal of this experiment is to study and measure
the horizontal handoff delay between two WLAN subnets as well as
detection delay of the prototype system,
and to evaluate the ability of the prototype system to handle such 
handoffs.   

To measure the impact of handoffs between two WLAN subnets, 
we used two access points (WP-II E made by Lucent Technologies) 
in our testbed and each AP was connected to an IP subnet,
with the beacon frequency of APs being configured as $1$ second.
A MH running Window XP 
was configured to have access to two IP subnets via a DHCP server. 
The MH was located in the overlapping coverage area of both APs, 
with one AP being configured as the preferred one and another as the backup. 

 \begin{figure}[htb]
	\centerline{\psfig{file=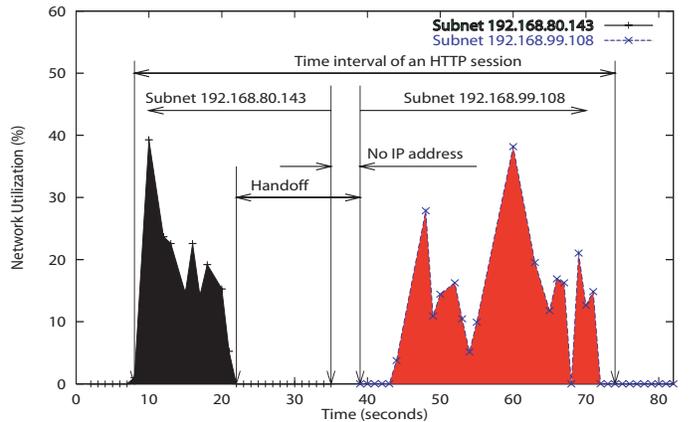,height=2.2in,width=3.5in}}
         \caption{Handoff Between WLAN subnets}
\label{fig:pre0} 
\end{figure}

In this experiment, we 
initiated a long-lived HTTP session 
(downloading a $23$M powerpoint file) via the preferred 
AP, and then triggered a handoff by turning off the AP's power.
This forced 
the MH to change its point of attachment from the 
preferred AP to the backup AP. 

It stands to reason that
data packets traveling through the preferred AP 
should immediately cut off when its power was off, 
marking a starting point for a network handoff (see Fig(\ref{fig:pre0})).
Such an observation would lead to a decomposition
of delay into two parts: network delay and 
detection delay.

\begin{table}[htbp]
\begin{center}
\caption{Performance Comparison}
\label{tab:wlanhandoff}
\begin{tabular}{||l|l|l|l||} \hline \hline
\multicolumn{4}{||c||} {\bf Overall download time} \\ \hline \hline 
\multicolumn{2}{||c} {\em No Handoff}&
\multicolumn{2}{|c||} {\em One Handoff} \\  \hline 
mean & stdev &mean &  stdev \\ \hline \hline
$41.3s$ & $6.34s$ &
$67.0s$ & $3.25s$ \\ \hline \hline
\multicolumn{4}{||c||} {\bf Network handoff delay} \\ \hline \hline
\multicolumn{2}{||c} {\em No Handoff}&
\multicolumn{2}{|c||} {\em One Handoff} \\  \hline 
mean & stdev &mean &  stdev \\ \hline \hline
$-$ & $-$ &
$17.4s$ & $3.4s$ \\ \hline \hline
\multicolumn{4}{||c||} {\bf Detection delay} \\ \hline \hline
\multicolumn{2}{||c} {\em No Handoff}&
\multicolumn{2}{|c||} {\em One Handoff} \\  \hline 
mean & stdev &mean &  stdev \\ \hline \hline
$-$ & $-$ &
$2.2s$ & $0.42s$ \\ \hline \hline
\end{tabular}
\end{center}
\end{table}

A horizontal handoff delay is defined as 
the time interval that starts when the rate of data packet 
associated with the preferred AP (one IP subnet) drops to zero 
and ends when the MH has attached to the backup IP subnet. 
A detection delay, which is closely related to our implementation, 
is defined as the time interval that starts when 
the MH has attached to backup subnet and ends when the MH starts
receiving and transmitting data packet via the new IP subnet. 
By examining the trace of IP packets, 
we were able to accurately measure both 
the horizontal handoff delay and 
the detection delay.

Table~\ref{tab:wlanhandoff} summarized the performance
measurements of the prototype system over $10$ 
independent runs. These results indicated that a handoff across different subnets 
could cause a $65\%$ additional delay in downloading a $23$M file.
It can be seen from Table~\ref{tab:wlanhandoff} and in Fig(\ref{fig:pre0}) that 
horizontal handoff delay, which includes the amount 
of elapsed time to release the 
IP address of the preferred subnet and 
to acquire the IP address of the backup subnet,
could take up to $17.4$ seconds on average, in which 
the absence of IP address can last for more than
$5$ seconds. The average detection delay was about
$2$ seconds over $10$ independent runs, with the standard deviation of $0.42s$ 
(see Table~\ref{tab:wlanhandoff} for details). 
These experimental results
highlight the importance of 
implementing a resilient system in order to survive long handoffs in a practical setting.  
 
\section{Conclusion and Future Work}
The real challenge in vehicular environments is 
how to best utilize geographically dispersed WLAN networks and 
omnipresent cellular networks such as CDMA, particularly 
in overlapping coverage areas,  
and how to provide a transparent and efficient failure 
recovery mechanism with 
the ability to adapt dynamically changing, inherently 
heterogeneous network environments.  

In this paper, 
we identify three problems that are critical to 
vehicular Internet access:  
(1) efficient HTTP failure recovery in network 
volatility while at the same time maintaining 
an user-perceived session continuity,
(2) network environment awareness, and (3) network adaptation 
via preemptive handoff in the presence of multiple networks
during HTTP sessions.   
We have designed and implemented a multi-tier prototype system   
with specifically tailored 
and carrier-agnostic features for vehicular Internet access. 

The main objective of this paper 
is to address these emerging problems stemming from 
inherently heterogeneous and dynamically changing vehicular 
environments, from the user's perspective, 
rather than from the wireless operator's perspective.
Therefore our focus is not placed on how to reduce handoff 
latency, but is rather placed on network awareness, 
network adaptation, and 
transparent HTTP failure recovery. This work differs 
significantly from 
prior research aiming at reducing handoff latency at 
layer $2$ and layer $3$.
It is worth noting that the prototype system is an 
application-layer approach that can fully exploit 
new advances at network and physical layers. 

The performance study showed 
that the prototype system can provide transparent and efficient 
HTTP failure 
recovery in the presence of vertical/horizontal/preemptive handoffs and 
is robust across transient network failures.
We showed that the phenomenon of the 
network hysteresis prevents MHs from taking advantage of 
available network resources in an overlapping coverage area of WLAN and 
a cellular network (CDMA) and presented an application-layer 
preemptive handoff to mitigate the impact of the network hysteresis
on the overall system throughput.  

The idea behind the proposed preemptive handoff is 
to initiate a network handoff based on network capacity, 
rather than on channel characteristics 
or signal strength level.
The advantage of
preemptive handoff lies in its ability to
dynamically select the best 
possible connection with a higher-bandwidth network in 
overlapping coverage areas,
which could result in a substantial reduction in elapsed time for 
long-lived HTTP sessions in a heterogeneous network consisting of
WLAN and CDMA/CDPD/GPRS. 

Many research issues remain for further exploration.  
Decreasing handoff latency is a pressing issue that 
needs to be adequately addressed. Our
future work will focus on how to reduce layer-two and 
layer-three handoff latency.    

\section{Appendix}

\noindent {\bf Proof:}  Let $T>0$ be a polling interval. Let  
$\{X_i, i\geq 1\}$ be the interarrival times of a Poisson process,
representing the time instants at which changes in network environment occur.  
It is obvious  that the random variables  
$X_i, i\geq 1$ are independently and exponentially distributed 
with the mean $\frac{1}{\lambda}$ \cite{Ross1996}. 
 
Define $S_0=0,~S_n=\sum\limits ^n _{i=1}X_i$. 
It follows that $S_i$ represents the arrival time of
the {\it i}th change event. Let

\begin{align} \label{equ:equ00}
n(T)=\sup \{n \geq 0:~S_n\leq T \}, 
\end{align}
where $n(T)$ is the Poisson counting process 
with rate $\lambda$, representing the number of
the event changes in the interval $(0,T]$.
Define detection delay, denoted by $D(.)$, as follows:\\

\begin{definition}
Detection delay at the time $t$ is 
\[ \mbox{D($t$)} = \left \{ 
\begin{array}{ll}
0, &  \mbox{if~polling~occurs~at~$t$}    \\
t-S_{n(t)}, & \mbox{otherwise} 
\end{array}
\right . \] \\
\end{definition} 

It can be seen from Fig(\ref{fig:proof})
that three event changes occur in the time interval $T$, 
the detection delay is thus determined by 
the time instant at which the last event change occurs and $T$, that is, 
$D(T)=T-S_{n(T)} = T-S_{3} $\footnote{$D_i = T-S_i$ refers to detection 
delay with respect to the {\it i}th event change 
in the time interval $T$}.  For any given interval $(0,t]$, 
the long-run mean average detection delay over the interval $(0,t]$ 
could be expressed as 
\begin{align}
E(D(\cdot))  & =  
\lim\limits_{t \rightarrow \infty} \frac{E(\int\limits_0^t D(\tau)d\tau)}{t} 
\nonumber  \\
 & =  
\lim\limits_{t\rightarrow \infty} 
\frac{E(\lfloor\frac{t}{T}\rfloor \int_0^T D(\tau)d\tau)}{t} 
\label{equ:delay}\\ 
& = 
\frac{E(\int_0^T (D(\tau)) d\tau)}{T}=\frac{\int_0^T E(D(\tau)) d\tau}{T} 
\label{equ:delay10},
\end{align}
where Eq(\ref{equ:delay}) follows the renewal reward theory \cite{Ross1996},
and Eq(\ref{equ:delay10}) assumes that integral and expectation operations are 
exchangeable.

Consider that the $n$th event occurs at time $s \in (0,T]$,
$S_n=\sum\limits ^n _{i=1}X_i$. It is obvious that 
$S_n$ follows the gamma distribution \cite{Ross1996} as follows:
\begin{align}\label{equ:gamma}
f_n(s) = \frac{ \lambda^n}{(n-1)!} s^{n-1}e^{-\lambda s},~s>0.
\end{align}

\begin{figure}[htb]
\psfig{file=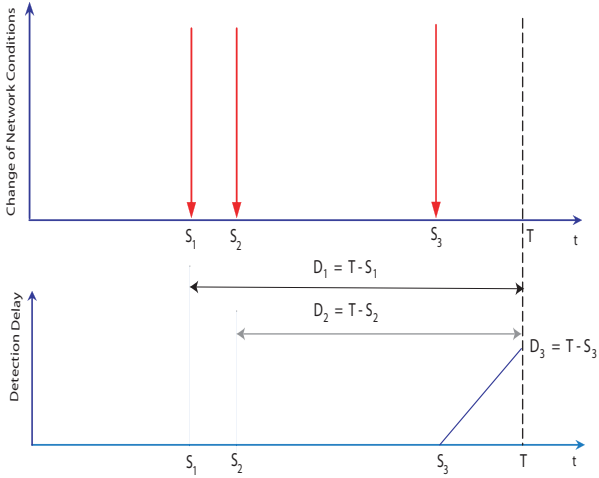,height=2.5in,width=3.1in}
\caption{Evolution of Detection Delay Function}
\label{fig:proof}
\end{figure}
Thus the expected detection delay at time $\tau \in (0,T]$ is written as 
\begin{align}\label{equ:eee}
& E(D(\tau))  =\int\limits_{0}^{\tau} (\tau-s) f_{n}(s) ds  \nonumber \\
&= \tau \left(1-\sum\limits_{i=0}^{n-1}\frac{(\lambda \tau)^i \exp(-\lambda \tau)}{i!}\right) \nonumber \\ 
&- \frac{n}{\lambda}\left(1-\sum\limits_{i=0}^{n} \frac{(\lambda \tau)^i\exp(-\lambda \tau)}{i!}\right). 
\end{align}

The analysis of general case could be hard since $n(\tau)$ is a 
random variable 
representing the number of change events in the interval $(0,\tau)$. 
Let's only consider the first event in $(0,T]$, which serves as 
a upper bound for general case.  
By inserting $n=1$ into Eq(\ref{equ:eee}) we obtain 

\begin{align}\label{equ:first}
E(D_1(\tau)) = \tau\left(1-\frac{1-\exp(-\lambda \tau)}{\lambda \tau}\right). 
\end{align}

\noindent Substituting Eq(\ref{equ:first}) into Eq(\ref{equ:delay1}) yields
\begin{align}\label{equ:delay1}
&E(D_1) = \frac{\int_0^T \tau(1-\frac{1-\exp(-\lambda \tau)}{\lambda \tau}) d\tau}{T} \nonumber \\
 &= \left(\frac{(T\lambda)^2 -2T\lambda +2 -2\exp(-T\lambda)}{2 T\lambda^2}\right),
\end{align}  
where $E(D_1)$ denotes the detection delay by only considering 
the first event occurring in the interval $[0,T]$.

Based on Eq(\ref{equ:equ00}), we have that $D_n = T-S_n \leq D_1 = T-S_1$ for $1\leq n$,
implying that $E(D_1) \leq E(D_n)$.  
The proof thus is completed. \hfill $\blacktriangle$

\section{Acknowledgment}
The authors would like to thank the anonymous reviewers 
for their critically reviewing the manuscript and 
for their truly helpful constructive comments. 
Yibei Ling would like to thank Dr. Shu-Chan Hsu in the 
Department of Cell Biology and
Neuroscience at Rutgers University for her 
encouragement and support. 
The preliminary material in this paper was presented in
part at the 2004 IEEE International Conference on Networking 
Sensing and Control.
\bibliography{seamless}

\begin{thebibliography}{10}

\bibitem{WAP2000}
Wireless {A}pplication {P}rotocol {W}hite {P}aper.
\newblock In {\em http://www.wapforum.org/what/WAP\_white\_pages.pdf}. WAP
  Forum, June 20 2000.

\bibitem{tarantella2005}
Introducing {T}arantella.
\newblock In {\em
  http://www.tarantella.com/support/documentation/enterprise/e3.3/help/en-us/b%
ase/gettingstarted/tarantella\_intro.html}. Sun Microsystem, 2005.

\bibitem{Anjum1999}
Farooq Anjum and Ravi Jain.
\newblock Performance of {TCP} over {L}ossy {U}pstream and {D}ownstream {L}inks
  with {L}ink {L}evel {R}etransmission.
\newblock In {\em Proceedings of the USENIX 1993 Winter Conference}, January
  1999.

\bibitem{Baker1996}
M.~Baker, X.~Zhao, S.~Cheshire, and J.~Stone.
\newblock Supporting mobility in mosquitonet.
\newblock In {\em Proc. of the 1996 USENIX Conference}, January 1996.

\bibitem{Caceres1995}
Ramon Caceres and Liviu Ifode.
\newblock Improving the {P}erformance of {R}eliable {T}ransport {P}rotocols in
  {M}obile {C}omputing {E}nvironments.
\newblock {\em IEEE Journal on Selected Areas in Communications},
  13(5):850--857, June 1995.

\bibitem{Cheshire1995}
S.~Cheshire and M.~Baker.
\newblock Experiences with a wireless network in mosquitonet.
\newblock In {\em Proc. of the 1995 IEEE Hot Interconnects Symposium}, August
  1995.

\bibitem{Claffy1998}
K.~Claffy, G.~Miller, and K.~Thompson.
\newblock The {N}ature of the {B}east: {R}ecent {T}raffic {M}easurements from
  an {I}nternet {B}ackbone.
\newblock In {\em Proceedings of INER's 1998 Conference}, July 1998.

\bibitem{Micrsoft2002}
Microsoft Corporation.
\newblock {IEEE} 802.11 {N}etwork {A}dapter {D}esign {G}uidances for {W}indows
  {XP}.
\newblock {\em Microsoft Corporation}, December 2002.

\bibitem{Katayama2001}
Minoru~Katayama et.al.
\newblock A method of achieving service continuity between different networks.
\newblock {\em IEICE Transactions on Communications (in Japanese)},
  J84-B(3):452--460, 2001.

\bibitem{Fikouras99}
N.A. Fikouras, K.~EI Malki, S.~R. Cvetkovic, and C.~Smythe.
\newblock Performance {E}valuation of {TCP} {O}ver {M}obile {IP}.
\newblock In {\em Proc. PIMRC}, 1999.

\bibitem{Anne1999}
Anne Fladenmuller and Ranil~De Silva.
\newblock The {E}ffect of {M}obile {IP} {H}andoff on the {P}erformance of
  {TCP}.
\newblock {\em Mobile Networks and Application}, 4(2):131--135, May 1999.

\bibitem{Stathes2002}
Stathes Hadjiefthymiades, Stamatis Papayiannis, and Lazaros Merakos.
\newblock Using path prediction to improve tcp performance in wireless/mobile
  environments.
\newblock {\em IEEE Coomunications Magazine}, 40(8):54--61, August 2002.

\bibitem{karagiannis-mobility}
Georgios Karagiannis and Geert Heijenk.
\newblock Mobility {S}upport for {U}biquitous {I}nternet {A}ccess.
\newblock In {\em ERICSSON Open Report}.
  http://www.ub.utwente.nl/webdocs/ctit/1/00000038.pdf, 2000.

\bibitem{Lewis1996}
Bil Lewis and Daniel~J. Berg.
\newblock {\em Threads {P}rimer: {A} {G}uide to {M}ultithreaded {P}rogramming}.
\newblock SunSoft Press, New York, 1996.

\bibitem{YiBingLin2002}
Yi-Bing Lin, Per-Chun Lee, and I.~Chlamtac.
\newblock Dynamic {P}eriodic {L}ocation {A}rea {U}pdate in {M}obile {N}etworks.
\newblock {\em IEEE Trans. on Veh. Technol}, 51(6):1494--1501, 2002.

\bibitem{Ling2004}
Yibei Ling, Wai Chen, Russell Hsing, and Onur Altinta.
\newblock Network {A}wareness and {A}daptation.
\newblock In {\em IEEE International Conference on Networking, Sensing and
  Control}, March 2004.

\bibitem{Ling2000}
Yibei Ling, Tracy Mullen, and Xiaola Lin.
\newblock Analysis of {O}ptimal {T}hread {P}ool {S}ize.
\newblock {\em ACM Operating System Review}, 34(2):42--55, 2000.

\bibitem{Ylianttila2001}
M.Ylianttila, M.~Pande, J.~Mäkelä, and P.~Mähönen.
\newblock Optimization {S}cheme for {M}obile {U}sers {P}erforming {V}ertical
  {H}andoffs between {IEEE} 802.11 and {GPRS/EDGE} {N}etworks.
\newblock In {\em Proceedings of IEEE Global Telecommunications Conference},
  volume~6, Auguest 2001.

\bibitem{Fikouras1999}
Fikouras N, El~Malki K, Cvetkovic SR, and Smythe C.
\newblock Performance of {TCP} and {UDP} during {M}obile {IP} {H}andoffs in
  {S}ingle-agent {S}ubnetworks.
\newblock In {\em Proc. IEEE Wireless Communications and Networking
  Conference}, 1999.

\bibitem{Okoshi1999}
Tadasho Okoshi, Masahiro Mochizuki, Yoshito, and Hideyuki Tokuda.
\newblock {M}obilesocket: {T}oward {C}ontinuous {O}peration for {J}ava
  {A}pplication.
\newblock In {\em IEEE 8th International Conference on Computer Communications
  and Networks}, 1999.

\bibitem{Parsa1999}
Christina Parsa and J.J. Garcia-Luna-Aceves.
\newblock Improving {TCP} {P}erformance over {W}ireless {N}etworks at the
  {L}ink {L}ayer.
\newblock In {\em Proceedings of 7th International Conference on Network
  Protocols}, 1999.

\bibitem{Perkins98}
Charles~E. Perkins.
\newblock {\em Mobile {N}etworking {T}hrough {M}obile {IP}}.
\newblock Sun Microsystems, New York, 1998.

\bibitem{Perkins99}
Charles~E. Perkins and Kuang-Yeh Wang.
\newblock Optimized {S}mooth {H}andoffs in {M}obile {IP}.
\newblock In {\em Proceedings of the The Fourth IEEE Symposium on Computers and
  Communications}, 1999.

\bibitem{Ross1996}
Sheldon~M. Ross.
\newblock {\em Stochastic {P}rocesses}.
\newblock John Wiley \& Sons, Inc., New York, 1996.

\bibitem{Henning2000}
Henning Schulzrinne.
\newblock {SIP} for {M}obility {A}pplication.
\newblock http://www.cs.columbia.edu/~hgs/sip/talks/von0006\_schulzrinne2.pdf,
  June 20 2000.

\bibitem{stemm98vertical}
Mark Stemm and Randy~H. Katz.
\newblock Vertical {H}andoffs in {W}ireless {O}verlay {N}etworks.
\newblock {\em Mobile Networks and Applications}, 3(4):335--350, 1998.

\bibitem{Weldlund99}
Elin Wedlun and Henning Schulzrinne.
\newblock Mobility {S}upport using {SIP}.
\newblock In {\em The Second ACM/IEEE International Conference on Wireless and
  Mobile Multimedia}, August 1999.

\end{thebibliography}
\bibliographystyle{plain}

\end{document}